\documentclass[aps,nofootinbib,floatfix,showpacs,preprintnumbers,twocolumn]{revtex4} %eqsecnum,
\usepackage{graphicx}
\usepackage{bm}
\usepackage{mathrsfs}
\usepackage{float}
\usepackage{tabularx} 
\usepackage{amsmath}

\input epsf

\graphicspath{{/Users/lstrigar/Dropbox/NeutrinoBackground/sterile/figures/}}
\def\lsim{\mathrel{\raise.3ex\hbox{$<$\kern-.75em\lower1ex\hbox{$\sim$}}}}
\def\gsim{\mathrel{\raise.3ex\hbox{$>$\kern-.75em\lower1ex\hbox{$\sim$}}}}

\def\cmm2{{\,\rm cm^{-2}}}
\def\cm2{{\,{\rm cm}^2}}
\def\cmm3{{\,{\rm cm}^{-3}}}
\def\gcmm3{{\,{\rm g\,cm^{-3}}}}

\def\R{{\cal R}}

\def\fun#1#2{\lower3.6pt\vbox{\baselineskip0pt\lineskip.9pt
  \ialign{$\mathsurround=0pt#1\hfil##\hfil$\crcr#2\crcr\sim\crcr}}}

\def\be{\begin{equation}}
\def\ee{\end{equation}}
\def\bea{\begin{eqnarray}}
\def\eea{\end{eqnarray}}

\begin{document}
%\modulolinenumbers[5]
%\linenumbers
%\preprint{FERMILAB-PUB-09-346-A-PPD, MAN/HEP/2009/31}

%\vspace{.2in}
\title{Solar neutrino physics with low-threshold dark matter detectors}

\author{J.~Billard}\email{billard@mit.edu} \affiliation{Department of Physics, Massachusetts Institute of Technology, Cambridge, MA 02139, USA}
\author{L.~Strigari}\thanks{Present Address: Mitchell Institute for Fundamental Physics and Astronomy, 
Department of Physics and Astronomy, Texas A \& M University, College Station, TX 77843-4242, USA} 
\email{strigari@physics.tamu.edu}  \affiliation{Department of Physics, Indiana University, Bloomington, IN 47405-7105, USA}
\author{E. Figueroa-Feliciano} \affiliation{Department of Physics, Massachusetts Institute of Technology, Cambridge, MA 02139, USA}

\date{\today}
\smallskip
\begin{abstract}
Dark matter detectors will soon be sensitive to Solar neutrinos via two distinct channels: coherent neutrino-nucleus scattering and neutrino electron elastic scattering. We establish an analysis method for extracting Solar model properties and neutrino properties from these measurements, including the possible effects of sterile neutrinos which have been hinted at by some reactor experiments and cosmological measurements. Even including sterile neutrinos, through the coherent scattering channel a 1 ton-year exposure with a low-threshold Germanium detector could improve on the current measurement of the normalization of the $^8$B Solar neutrino flux down to 3\% or less. Combining with the elastic scattering data will provide constraints on both the high and low energy survival probability, and will improve on the uncertainty on the active-to-sterile mixing angle by a factor of two. This sensitivity to active-to-sterile transitions is  competitive and complementary to forthcoming dedicated short baseline sterile neutrino searches with nuclear decays.

\end{abstract}
\maketitle

\section{Introduction}
\label{sec:intro}
\par Dark matter detectors are rapidly improving sensitivity~\cite{experiments}, and as they continue to increase in size and reduce thresholds, they will encounter the neutrino background, at which point Solar, atmospheric, and diffuse supernova neutrinos will interfere with a potential dark matter signal~\cite{neutrinofloor}. Neutrino interactions in these detectors will occur through both coherent neutrino-nucleus scattering (CNS)~\cite{CNS} and neutrino-electron elastic scattering (ES). Understanding the expected neutrino signals will be crucial not only for the purposes of extracting a dark matter signal, but also for extracting properties of neutrinos~\cite{properties,Baudis:2013qla} and their astrophysical sources. 

\par Focusing in particular on Solar neutrinos, experimental measurements have provided a wealth of information on fundamental properties of neutrinos and on properties of the Sun (for recent reviews see Refs.~\cite{reviews}). Through these measurements, it is now well-established that the transformation of high energy neutrinos from the Sun is due to the matter-induced MSW effect, which provides the explanation for the detected electron neutrino event rate on Earth relative to the predicted rate. Neutrino mass differences and mixing angles are then determined by combining Solar data with data from atmospheric, accelerator, and reactor neutrino experiments~\cite{deGouvea:2013onf}. 

\par Solar neutrino data also can provide an important test of Standard Solar Models (SSMs). Recent 3D rotational hydrodynamical simulations~\cite{Asplund:2009fu} suggest a lower abundance of metals in the Solar core relative to previous models~\cite{Grevesse:1998bj}, which implies a reduced temperature in the Solar core and a corresponding reduction in some of the neutrino fluxes. Though helioseismology data are inconsistent with a lower metallicity, future measurements of neutrino fluxes may be able to distinguish between a high or low metallicity Solar model. 

\par In addition to providing a test of SSMs, Solar neutrinos may also provide a probe of exotic new physics. In particular, some reported measurements appear inconsistent with the standard picture of neutrino mass differences and mixing angles. First, there is a deficit of electron neutrinos measured~\cite{Giunti:2006bj,Giunti:2010zu} in the radioactive source experiments of the GALLEX~\cite{Kaether:2010ag} and SAGE~\cite{Abdurashitov:2009tn} Solar neutrino detectors. Second, very short baseline (VSBL) neutrino experiments with distances of $< 100$ m indicate a deficit of electron anti-neutrinos (the reactor neutrino anomaly)~\cite{Mention:2011rk}. Both of these results can be explained by an additional neutrino with a mass splitting $\Delta m^2 \sim 1$ eV$^2$. Additional possible evidence for sterile neutrinos comes from short-baseline experiments (LSND and MiniBooNE)~\cite{Aguilar:2001ty,AguilarArevalo:2010wv,Aguilar-Arevalo:2013pmq}. Cosmological measurements may also be interpreted as favoring the existence of light sterile neutrinos~\cite{cosmology}.  Light sterile neutrinos can also be searched for using both long baseline reactors and Solar neutrino experiments~\cite{Giunti:2009xz,Palazzo:2011rj,Palazzo:2012yf} (For a recent general review on sterile neutrinos see Ref.~\cite{Kopp:2013vaa}).  

\par There are additional possible hints for sterile neutrinos that come directly from Solar neutrinos. For example, measurements of the Solar ${}^8$B electron neutrino flux by the Sudbury Neutrino Observatory (SNO)~\cite{Aharmim:2011vm}, Super-Kamiokande (SK)~\cite{Abe:2010hy}, and Borexino~\cite{Bellini:2008mr}, combined with the SNO neutral current (NC) measurement, indicate a constant electron neutrino survival probability over the ${}^8$B energy range. In contrast, the LMA-MSW solution predicts that at the lowest energies that SNO and SK are sensitive to, there is an upturn in the survival probability coming from the fact that at such energies the flavor transformations are dominated by vacuum effects. New physics in the neutrino sector, such as non-standard neutrino interactions~\cite{Friedland:2004pp} or transitions into a non-active sterile component~\cite{deHolanda:2010am}, can predict an energy-independent survival probability in this intermediate regime. 

\par Motivated by the prospects for improving understanding the SSM and neutrino properties, in this paper we perform a general study of the sensitivity of dark matter detectors to Solar neutrinos. We include the possibility of sterile neutrinos in our analysis within a specific theoretical framework involving a single new sterile neutrino with mass splitting of $\Delta m^2 \sim$ eV$^2$. We discuss the utility of both CNS and ES data from a dark matter detector. Our primary results show that CNS data substantially improve the measurement of the normalization of the ${}^8$B Solar neutrino flux, and the ES data  substantially improve the measurement of the neutrino mixing parameters. Interestingly, combining these two independent channels together can lead to much improved constraints on the active-to-sterile mixing angle.

\par This paper is organized as follows. In Section~\ref{sec:cns} we briefly review the physics of both coherent neutrino scattering and neutrino-electron scattering, and discuss detection prospects for Solar neutrinos through CNS and ES. In Section~\ref{sec:models} we briefly discuss a 3+1 model with a single new sterile neutrino. In Section~\ref{sec:mcmc} we introduce our methodology for constraining the parameters of the 3+1 sterile neutrino model with CNS and ES data from a dark matter detector. In Section~\ref{sec:results} we present the results of our analysis, and then close in Section~\ref{sec:conclusion} with our discussion and conclusions. 

\section{Extracting coherent neutrino scattering and elastic scattering signals}
\label{sec:cns} 

\par In this section we briefly review the coherent neutrino and neutrino electron scattering processes. We then discuss the properties of future dark matter detectors that will be sensitive to both CNS through nuclear recoils and neutrino-electron scattering through electron recoils. 

\par It has been shown by Freedman \cite{freedman} that the neutrino-nucleon elastic interaction leads to a coherence effect implying a neutrino-nucleus cross section that approximately scales as the atomic number ($A$) squared when the momentum transfer is below a few keV. At tree level, the neutrino-nucleon elastic scattering proceeds through the exchange of a $Z$ boson within a neutral current interaction. The resulting differential neutrino-nucleus cross section as a function of the recoil energy $T_R$ and the neutrino energy $E_\nu$ is~\cite{freedman2}
\begin{equation}
\frac{d\sigma_{CNS}(E_\nu, T_R)}{dT_R} = \frac{G^2_f}{4\pi}Q^2_w m_N \left(1 - \frac{m_NT_R}{2E^2_{\nu}}  \right)F^2(T_R),
\end{equation}
where $m_N$ is the target nucleus mass, $G_f$ is the Fermi coupling constant, $Q_w = N - (1-4\sin^2\theta_w)Z$ is the weak nuclear hypercharge with $N$ the number of neutrons, $Z$ the number of protons, and $\theta_w$ the weak mixing angle.  $F(T_R)$ is the nuclear form factor that describes the loss of coherence for recoil energies above $\sim$10~keV. In the following, we will consider the standard Helm form factor~\cite{lewin}. 

\par Future dark matter detectors will also soon be sensitive to the neutrino-electron electroweak interaction. This proceeds through the exchange of a $Z$ boson (neutral current) and the exchange of a $W$ boson (charged current). The latter is only possible in the case of an incoming $\nu_e$. The resulting cross section is~\cite{Marciano:2003eq,Formaggio:2013kya} 
\begin{align}
\frac{d\sigma_{ES}(E_\nu, T_r)}{dT_r} = & \frac{G_f^2m_e}{2\pi}\left[ (g_v + g_a)^2 \right. \nonumber \\ 
&\left. + (g_v - g_a)^2\left(1 - \frac{T_r}{E_\nu}\right)^2 + (g_a^2 - g_v^2)\frac{m_e T_r}{E_\nu^2}  \right], 
\end{align}
where $m_e$ is the electron mass, $g_v$ and $g_a$ are the vectorial and axial coupling respectively and are defined such that
\begin{equation}
g_v = 2\sin^2\theta_w - \frac{1}{2} \ \ \ \ \ \ \ g_a = - \frac{1}{2}. 
\end{equation}
In the particular case $\nu_e + e \rightarrow \nu_e + e$, the interference due to the additional charged current  contribution implies a shift in the vectorial and axial coupling constants such that $g_{v,a} \rightarrow g_{v,a} + 1$. Due to the rather large difference in the $\nu_e + e$ and $\nu_{\mu,\tau}+e$ cross sections of almost an order of magnitude, by measuring the neutrino-electron scattering rate, one can derive the neutrino electron survival probability. The standard MSW-LMA solution leads to a rather flat neutrino-electron survival probability below 1 MeV of about 0.545~\cite{Friedland:2004pp}.

\par
\begin{figure}%[htp]
\begin{center}
\includegraphics[width=0.49\textwidth]{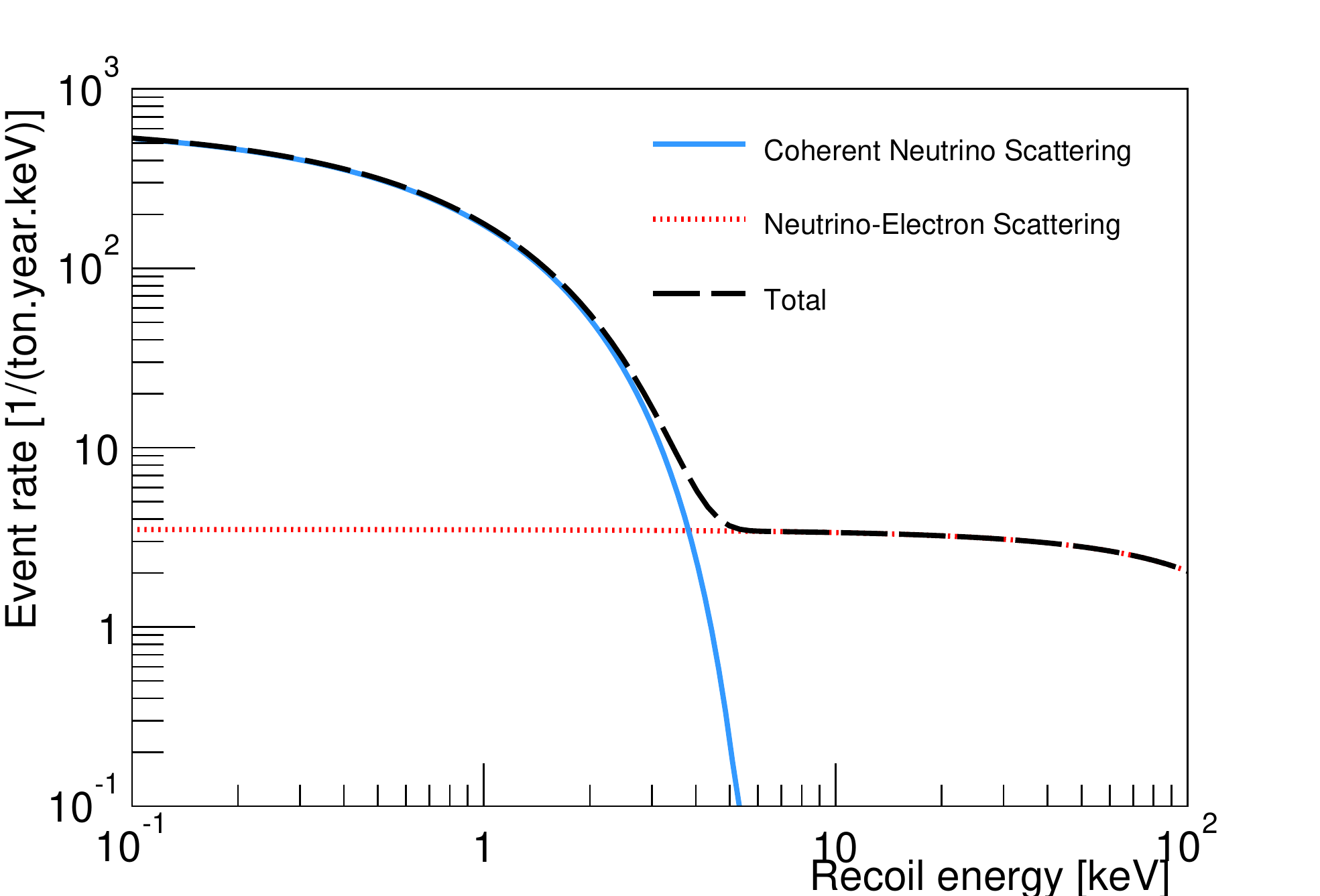}
\end{center}
\caption{Neutrino induced backgrounds in a low-threshold Ge dark matter detector. The $^8$B induced nuclear recoils (CNS) and the $pp$ induced electronic recoils (ES) are shown as the blue solid and red dashed lines respectively. These event rates have been computed using the high metallicity standard solar model, $P_{ee}$ = 0.55 for {\it pp} neutrinos and $P_{es} = 0$ at all neutrino energies.
\label{fig:spectrum}
}
\end{figure}

 Figure~\ref{fig:spectrum} shows the event rate spectra from $^8$B induced CNS nuclear recoils (blue solid line) and $pp$ induced ES electronic recoils (red dashed line) as a function of recoil energy. The former neutrinos are produced from the reaction ${}^8B \rightarrow {}^8Be + e^+ + \nu_e$ and the latter are produced from $p + p \rightarrow {}^2H + + e^+ + \nu_e$. We plot the rate above a recoil energy threshold of 0.1~keV for a Ge detector. With a 0.1~keV energy threshold, we are sensitive to most $pp$ neutrinos in the ES channel and  to neutrino energies above approximately 1.9 MeV in the CNS channel. In such configurations, both channels are almost perfectly pure samples of $pp$ and $^8$B neutrinos which then offer the unique possibility to accurately probe the solar neutrino physics in both the vaccum and the matter dominated regimes with a single experiment. As a matter of fact, with a one ton-year exposure Ge detector,  one expects about $\sim$ 500 neutrino events in both the CNS and ES channels above 0.1 keV recoil energy.

\par Several Dark Matter detection techniques for lowering the experimental threshold are under development. For cryogenic crystal experiments, the use of high electric field across the crystals results in a significant amplification of the total phonon signal~\cite{Luke,Neganov}, with the potential to significantly lower the threshold. The SuperCDMS collaboration has shown the possibility to lower the threshold down 170~eVee (electron equivalent) which is equivalent to a threshold on the nuclear recoil energy of about 800~eV, with lower thresholds projected in the future~\cite{Agnese:2013jaa}. Using CaWO$_{4}$ cryogenic crystals, the CRESST collaboration recently demonstrated a nuclear recoil threshold of 600~eV~\cite{Angloher:2014}.  Another possibility is the use of the secondary scintillation signal (S2) in Xe experiments as demonstrated by the XENON10 collaboration~\cite{Angle:2011}, where they performed an S2-only analysis with a threshold of 5 electrons, corresponding to 1.4~keV nuclear recoil energy. 

\par Since in this paper we are trying to evaluate the physics that may be achieved with future low-threshold dark matter detectors, we will assume an experimental threshold of 0.1~keV and will not consider additional sources of background and no detection of dark matter particles.
Due to the very different spectral shapes of the CNS and ES signal (see Fig.~\ref{fig:spectrum}), the discrimination power between these two populations of events is large enough that it does not induce additional systematics in the neutrino parameter estimations. Therefore, event identification between ES and CNS is not assumed, although substantial discrimination power between electron and nuclear recoils can be achieved by Dark Matter experiments using ionization or light yield quantities (typically at the expense of a higher analysis threshold). 

\par Note that for all the calculations in Figure~\ref{fig:spectrum} and for the following results we utilize a Ge target, although our quantitative results will not change substantially for different targets. As a matter of fact, the lighter is the target nucleus, the easier it is to detect CNS events from $^8$B neutrinos as the required energy threshold increases: 4~keV (Xe), 7.9~keV (Ge), 20~keV (Si), and 35~keV (CaWO$_{4}$ thanks to the light O target). However, CNS is a coherent process that scales as $A^2$ implying larger event rates for heavier targets at a fixed exposure. For example, with a 0.1~keV threshold, the CNS rate for a Xe target is about a factor of two larger than for a Ge target for a similar exposure. From a practical perspective, it is likely that a Ge target will be able to more easily achieve the low thresholds that we discuss relative to a Xe target. However, for larger thresholds Xe targets are more likely to achieve the exposures that we consider below. In the case of the neutrino-electron scattering, we checked that the event rate is fairly insensitive to the particular choice of target nucleus.

%%%%%% Figure: Spectrum %%%%%%%%%
%%%%%%%%%%%%%%%%%%%%%%%%

\section{$3+1$ neutrino model}

\label{sec:models}
\par In this section, we move on to discuss the theoretical model that we use for neutrino oscillations. Within this model-dependent framework, our goal is to then determine in section~\ref{sec:mcmc} what CNS and ES measurements from a dark matter detector would add to the existing measurements from reactors and other Solar neutrino experiments. For simplicity, we focus on the theoretical model with one new mass splitting that is due to a single sterile neutrino that is much larger than the measured mass splittings $|\Delta m_{21}^2|$ and $|\Delta m_{32}^2|$. This model can be extended to also include more than one additional sterile neutrino, see e.g.~Ref.~\cite{Giunti:2009xz}. Here we simply review the formulae that are required to calculate transition probabilities for this model with one additional sterile neutrino; for a more complete discussion of this model see Ref.~\cite{Palazzo:2011vg}.
 
\par With one additional sterile neutrino, there are a total of 6 angles that are required to describe the neutrino mixing matrix, $\theta_{12}$, $\theta_{13}$, $\theta_{23}$, $\theta_{14}$, $\theta_{24}$, and $\theta_{34}$. For the analysis in this paper we will take $\theta_{24} = \theta_{34} = 0$, so that the only possible new non-zero angle is $\theta_{14}$. Small values of $\theta_{24}$ and $\theta_{34}$ are deduced from the results of reactor experiments~\cite{Kopp:2013vaa}, so setting these ``non-solar'' angles to zero will not affect the results that we present hereafter. If we were to consider nonzero values of $\theta_{24}$ and $\theta_{34}$, we would have to also account for the possibility of additional small CP violating phases on top of the one in the standard three-neutrino model. 

\par For our assumption of $\theta_{24}$ and $\theta_{34}$, the relevant elements of the mixing matrix that determine mixing between the electron flavor and the mass eigenstates are~\cite{Giunti:2009xz,Palazzo:2011vg}
\begin{eqnarray}
U_{e1} &=& c_{14} c_{13} c_{12}  \\
U_{e2} &=& c_{14} c_{13} s_{12} \\
U_{e3} &=& c_{14} s_{13}  \\
U_{e4} &=& s_{14}
\end{eqnarray} 
where $s_{\imath \jmath} = \sin \theta_{\imath \jmath}$ and $c_{\imath \jmath} = \cos \theta_{\imath \jmath}$. The mixing between the sterile component and the mass eigenstates are controlled by 
\begin{eqnarray}
U_{s1} &=& -s_{14} c_{13} c_{12}  \\
U_{s2} &=& -s_{14} c_{13} s_{12} \\
U_{s3} &=& -s_{14} s_{13}  \\
U_{s4} &=& c_{14}
\end{eqnarray} 
In addition to the mixing elements in vacuum, we will also need the effective mixing matrix elements in matter at the electron neutrino production point. These are given by 
\begin{eqnarray}
U_{e1}^m &=& c_{14} c_{13} c_{12}^m  \\
U_{e2}^m &=& c_{14} c_{13} s_{12}^m \\
U_{e3}^m &=& U_{e3} = c_{14} s_{13}  \\
U_{e4}^m &=& U_{e4} = s_{14}. 
\end{eqnarray} 
In these equations the matter mixing angles are defined through 
\begin{eqnarray}
\frac{k_m}{k} \sin 2 \theta_{12}^m &=& \sin 2 \theta_{12} \\
\frac{k_m}{k} \cos 2 \theta_{12}^m &=& \cos 2 \theta_{12} - v_x \gamma^2 - v_x r_x \alpha^2 
\end{eqnarray} 
where $k$, $k_m$ are the neutrino wavenumbers in vacuum and in matter. The ratio of the neutral current to the charged current potential is $r_x = 0.25$, $\gamma = c_{13} c_{14}$, $\alpha = - s_{14} s_{13}$, and $v_x = V_{cc} / k$, and we take the matter potential to be $V_{cc} = 10^{-11}$ eV. Note that here we have not accounted for the small variation in the matter potential with radius in the Sun. 

\par With the above assumptions for the mixing matrix elements, the probability to detect an electron neutrino of flavor $\alpha = e, \mu, \tau, s$, where here $s$ stands for sterile, that is produced in the Sun is~\cite{Giunti:2009xz,Palazzo:2011vg}
\begin{equation}
P_{e \alpha} = \sum_{\imath = 1}^4 U_{\alpha \imath}^2 (U_{e\imath}^m)^2. 
\label{eq:p_model}
\end{equation} 
This probability does not account for phase information that gets lost by a spatial averaging over the neutrino production region and by smearing of energy. Note that we do not account for small Earth-induced matter oscillations for Solar neutrinos~\cite{Renshaw:2013dzu}. For the Solar neutrino analysis, there is no dependence on the mass splitting $\Delta m_{41}^2$, as oscillations due to this mass difference are averaged out over the Earth-Sun baseline. 

\par For a fixed $E_\nu$ we have a unitarity constraint 
\begin{equation} 
P_{ee} + P_{ea} + P_{es} = 1, 
\label{eq:unitarity} 
\end{equation}
where $P_{ea}$ is the probability that an electron neutrino transitions into a mu/tau neutrino component. 

\section{Data analysis} 
\label{sec:mcmc} 
\par With the above theoretical model in place, in this section we discuss our analysis of the data sets. We begin by discussing the analysis of the very long baseline KamLAND data, and then move on to discuss our analysis of the Solar neutrino data. For the latter analysis we highlight the new information that both CNS and ES data from a dark matter detector can provide on parameters of the 3+1 model.  Although we do not use recent measurements from Daya Bay, Reno, and Double Chooz of non-zero $\sin^2 \theta_{13}$ in our analysis, in the discussion section we estimate the implications that these short baseline reactor data have on our results. 

\subsection{Reactor data} 

\par In order to implement our analysis methods in this section, we need an expression for the neutrino survival probability in vacuum. With the assumptions in Section~\ref{sec:models}, for propagation in vacuum the electron neutrino survival probability is 
\begin{equation}
P_{ee} = 1 - \sum_{\imath < \jmath} 4 | U_{e \imath} |^2  | U_{e \jmath} |^2 \sin^2 \left( \frac{ \Delta m_{\imath \jmath}^2 L}{4 E_\nu} \right). 
\end{equation}

\par For the case of oscillations driven by the mass-squared difference $\Delta m_{21}^2$, as will be appropriate for the analysis of KamLAND data, the survival probability can be approximated as
\begin{equation}
P_{ee} = c_{14}^4 c_{13}^4 P_{ee}^{2\nu} + c_{14}^4 s_{13}^3 + s_{14}^4, 
\label{eq:vacuum}
\end{equation}
where the two flavor survival probability in vacuum is 
\begin{equation}
P_{ee}^{2\nu} = 1 - 4 s_{12}^2 c_{12}^2 \sin^2 \left[ \frac{\Delta m_{21}^2 L}{4 E} \right]. 
\label{eq:matter} 
\end{equation} 

\par  For KamLAND, we use the data and the prescription outlined in Ref.~\cite{Gando:2010aa}, which is appropriate for determining how small angles $\theta_{13}$ and $\theta_{14}$ affect the values of $\theta_{12}$ and $\Delta m_{21}^2$ that are determined within a two-flavor neutrino framework. In particular, KamLAND provides a measurement of the survival probability as a function of the following quantity, 
\begin{equation} 
x(E_\nu,L) \equiv \frac{1}{\sin^2 \hat \theta_{12}} \left \langle \sin^2 2\theta_{12 M} \sin^2 \left( \frac{ \Delta m_{12 M}^2 L}{4E_\nu} \right) \right \rangle, 
\label{eq:x_kamland}
\end{equation}
where the matter-modified angle and mass splitting is 
\begin{equation} 
\sin^2 2 \theta_{12 M} = \frac{\sin^2 2 \theta_{12}}{ (\cos 2 \theta_{12} - A/\Delta m_{21}^2)^2 + \sin^2 2 \theta_{12} }
\end{equation} 
and 
\begin{equation}
\Delta m_{21 M}^2 = \Delta m_{21}^2 \sqrt{ ( \cos 2 \theta_{12} - A/\Delta m_{21}^2 )^2 + \sin^2 2 \theta_{12} }. 
\end{equation}
Here the $A = - 2 \sqrt{2} G_F \tilde N_e E_\nu$, where $\tilde N_e = N_e \cos^2 \theta_{13}$ and $N_e \simeq 2 N_A$ g cm$^{-3}$ is the electron number density. In Equation~\ref{eq:x_kamland} the hat over the angles denotes the best fitting solution from a two flavor analysis, and the subscript $M$ accounts for matter oscillations. We take $\Delta m_{21}^2$ as its measured value from a two-flavor analysis, $\Delta m_{21}^2 = 7.5  \pm 0.2 \times 10^{-5}$ eV$^2$~\cite{Gando:2010aa}. With this choice we then calculate the vacuum survival probability in Equation~\ref{eq:vacuum} as a function of the three mixing angles.

\subsection{Solar data}  

\par For our Solar analysis, we use data from SNO, SK, Borexino, Homestake, and Gallium experiments.  SNO and SK are mostly sensitive to ${}^8$B neutrinos, with a small contribution from ${\it hep}$ neutrinos. For SK we use the ES energy spectrum over the electron recoil kinetic energy range [5.0-20] MeV~\cite{Abe:2010hy}. For SNO we use the total NC rate as determined from the three-phase analysis~\cite{Aharmim:2011vm}. For Borexino we use measurements of the ${}^7$Be~\cite{Bellini:2011rx} and {\it pep}~\cite{Collaboration:2011nga} neutrino fluxes. We also include the Borexino ES energy spectrum over electron recoil kinetic energy range [3.0-13.0] MeV~\cite{Bellini:2008mr}; though at high energies this data is much less sensitive than that of SK, we include it for completeness because it extends to lower energies than SK. For Homestake we use the final results from Ref.~\cite{Cleveland:1998nv}, and for Gallium we use the combined analysis of Ref.~\cite{Abdurashitov:2009tn}.

\begin{table*}
\caption{\label{tab:constraints} Experiments, observables, and parameters that are best constrained by each of the experiments that are utilized in our analysis.} 
\begin{ruledtabular}
\begin{tabular}{llcc}
  Experiment & Observable & Best constrained parameters & Reference\\
  	\hline
    SNO & Neutral Current rate & $f_{8B}$, $\sin^2 \theta_{12}$, $\sin^2 \theta_{14}$ &~\cite{Aharmim:2011vm}  \\
    SK & Elastic Scattering rate & $f_{8B}$, $\sin^2 \theta_{12}$, $\sin^2 \theta_{14}$ &~\cite{Abe:2010hy} \\
    Borexino & Elastic Scattering rate & $f_{7Be}$, $f_{pep}$, $f_{8B}$, $\sin^2 \theta_{12}$, $\sin^2 \theta_{14}$ &~\cite{Bellini:2008mr,Bellini:2011rx,Collaboration:2011nga}  \\
    Homestake & Integrated Capture rate & $f_{8B}$, $f_{7Be}$&~\cite{Cleveland:1998nv} \\
    Gallium & Integrated Capture rate & $f_{8B}$, $f_{7Be}$, $f_{pp}$ &~\cite{Abdurashitov:2009tn} \\
 KamLAND &$\bar \nu_e$ disappearance& $\Delta m_{21}^2$, $\sin^2 \theta_{12}$, $\sin^2 \theta_{13}$, $\sin^2 \theta_{14}$&~\cite{Gando:2010aa} \\
\end{tabular}
\end{ruledtabular}
\end{table*}

%%%%%%%%%%% Figure: MCMC w/o CNS %%%%%%%%%%
\begin{figure*}%[htp]
\begin{center}
\includegraphics[width=\textwidth]{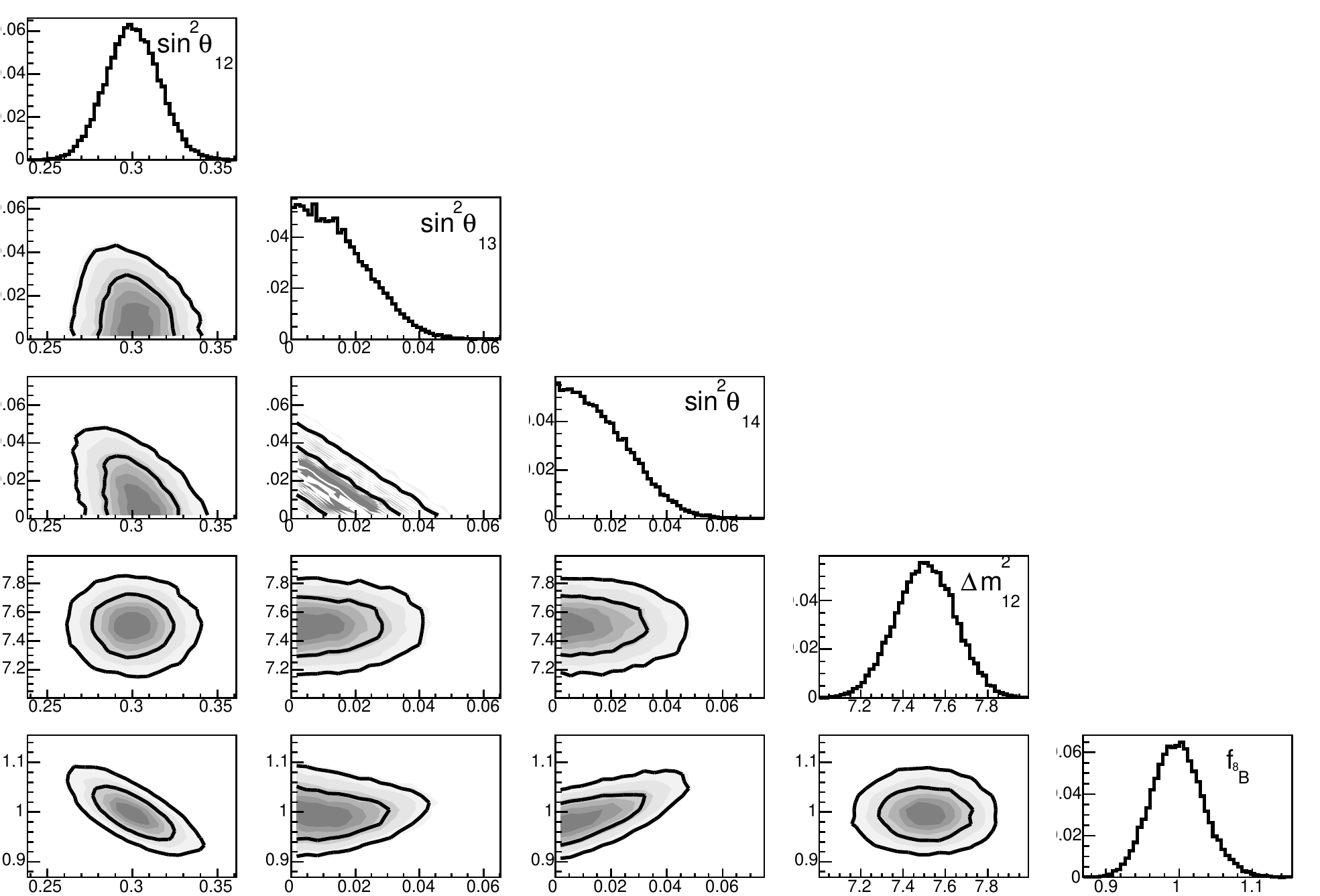}
\end{center}
\caption{Marginalized posterior probability density functions for selected model parameters from our MCMC analysis considering only existing the data from the experiments listed in Table~\ref{tab:constraints} and the high metallicity SSM~\cite{Grevesse:1998bj} listed in Table~\ref{tab:ssm}. Along the off diagonal are the correlations between the different parameters, where the thick contours reflect the 68\% and 95\% C.L. of the joint distributions. The other parameters $\{f_{^7Be}, f_{pep}, f_{pp}, f_{hep}, f_{CNO}\}$, not shown here, have been marginalized over.
\label{fig:mcmc}
}
\end{figure*}

\begin{table*}
\caption{\label{tab:params}  Constraints on parameters that we deduce from our MCMC analysis presented in Fig.~\ref{fig:mcmc}, compared to previous constraints on the parameters in Column 3 as determined from the reference indicated. The errors on the ``previous results'' for $\sin^2\theta_{13}$ are given in terms of the statistical plus systematic uncertainty. 
}
\begin{ruledtabular}
\begin{tabular}{cccc}
  Parameter & our result (68\% C.L.) & previous result & Reference\\
  	\hline
   $f_{8B}$ & 0.998$\pm$ 0.034 & $0.941\pm 0.036$ &~\cite{Aharmim:2011vm}\\
   $\sin^2 \theta_{12}$ & 0.300$\pm$0.016 &$0.307_{-0.015}^{+0.017}$ & \cite{Gando:2010aa}\\
   $\sin^2\theta_{13}$ & $<$0.030 (90\% C.L.) & $0.0235 \pm 0.0042 \pm 0.0013$ ,$ 0.0291 \pm 0.0035 \pm 0.0051$ & \cite{An:2012eh}, \cite{Ahn:2012nd}\\
    $\sin^2 \theta_{14}$ & $<$0.034 (90\% C.L.) & $< 0.04$ & \cite{Palazzo:2011rj}\\	
    $\Delta m_{21}^2$ ($\times10^{-5}$ eV$^2$) & 7.5$\pm$0.14 & 7.5$\pm$0.2 & \cite{Gando:2010aa}\\	
\end{tabular}
\end{ruledtabular}
\end{table*}

\par All of the solar experiments do not directly measure the electron neutrino survival probability, but rather the neutrino survival probability convolved with a cross section and the appropriate neutrino spectrum. For ES measurements, taking $P_{ea}$ in Equation~\ref{eq:p_model} as the appearance probability for mu and tau neutrinos, the prediction for the ES energy spectrum relative to the scenario in which there is no neutrino flavor transformations is 
\begin{widetext}
\begin{equation}
\R_{ES} (T_{\mathit{eff}}) = f_\imath \frac{\int f_\nu(E_\nu)\left[ P_{ee}(E_\nu) \frac{d\sigma_{ES,e}}{dT_e} +  P_{ea}(E_\nu) \frac{d\sigma_{ES,a}}{dT_e}\right]G(T_e,T_{\mathit{eff}}) dE_\nu dT_e}
{\int f_\nu(E_\nu) \frac{d\sigma_{ES,e}}{dT_e}  G(T_e,T_{\mathit{eff}}) dE_\nu dT_e}. 
\label{eq:es}
\end{equation} 
\end{widetext}

\begin{table}
\caption{\label{tab:ssm} Flux normalizations for the high metallicity GS98-SGII~\cite{Grevesse:1998bj} that are utilized in our analysis.} 
\begin{ruledtabular}
\begin{tabular}{llc}
  Neutrino flux & SSM prior & units\\
  	\hline
    $pp: p + p \rightarrow {}^2H + e^+ + \nu_e$ & $5.98 ( 1 \pm 0.006)$ & $10^{10}$ cm$^{-2}$ s$^{-1}$\\
    $pep: p + e^- \rightarrow {}^2H + \nu_e$ & $1.44 (1 \pm 0.012)$ & $10^{8}$ cm$^{-2}$ s$^{-1}$\\
    ${}^7Be: {}^7Be + e^- \rightarrow {}^7Li + \nu_e$ & $5.00 (1 \pm 0.07)$ & $10^{9}$ cm$^{-2}$ s$^{-1}$\\
    ${}^8B: {}^8B \rightarrow {}^8Be + e^+ + \nu_e$ & $5.58 (1 \pm 0.14)$ & $10^{6}$ cm$^{-2}$ s$^{-1}$\\
    $hep: {}^3He + p \rightarrow {}^4He + e^+ + \nu_e$ & $8.04 (1 \pm 0.30)$ & $10^{3}$ cm$^{-2}$ s$^{-1}$\\
    ${}^{13}C: {}^{13}N \rightarrow {}^{13}C + e^+ + \nu_e$ &  $2.96(1\pm 0.14)$ & $10^{8}$ cm$^{-2}$ s$^{-1}$\\
    ${}^{15}N: {}^{15}O \rightarrow {}^{15}N + e^+ + \nu_e$ &  $2.23(1\pm 0.15)$ & $10^{8}$ cm$^{-2}$ s$^{-1}$\\
    ${}^{17}O: {}^{17}F \rightarrow {}^{17}O + e^+ + \nu_e$&  $5.52(1\pm 0.17)$ & $10^{6}$ cm$^{-2}$ s$^{-1}$\\
\end{tabular}
\end{ruledtabular}
\end{table}

In this equation, $f_\nu (E_\nu)$ is the unit-normalized neutrino energy spectrum, and $f_\imath$ is the ratio of the full neutrino flux of the $\imath^{th}$ component relative to a Standard Solar Model (SSM) prediction, with $\imath = {}^8B, {}^7Be, CNO, pep, pp, hep$.
The electron neutrino elastic scattering cross section is $d\sigma_{ES,e}/dT_e$ and $d\sigma_{ES,a}/dT_e$ is the mu and tau neutrino elastic scattering cross section. These cross sections are functions of the true recoil electron kinetic energy $T_e$. The function $G(T_e,T_{\mathit{eff}})$ is the gaussian energy response, which is a function of $T_e$ and measured electron kinetic energy $T_{\mathit{eff}}$. The SK and Borexino ES data sets that we utilize are in the form of an integrated number of events relative to the SSM prediction in each energy bin, so to compare to our predictions we simply integrate Equation~\ref{eq:es} over the appropriate $T_{\mathit{eff}}$ corresponding to the energy range covered by each bin, using the measured $G(T_e,T_{\mathit{eff}})$ for each experiment.

\par To these spectral measurements we add the SNO NC flux measurement of $5.25 \pm 0.20 \times 10^6$ cm$^{-2}$ s$^{-1}$, which is derived from the measured event rate above the deuterium breakup threshold of 2.2 MeV~\cite{Aharmim:2011vm}. 
For SNO, the rate relative to the SSM is then
\begin{equation}
{\cal R}_{NC} = f_{8B} \frac{\int [1 - P_{es}(E_\nu)]f(E_\nu)\frac{d\sigma_{\nu - d}}{d E_\nu}(E_\nu)dE_\nu}{\int f(E_\nu)\frac{d\sigma_{\nu - d}}{d E_\nu}(E_\nu)dE_\nu}, 
\label{eq:rnc} 
\end{equation}
where $ {d\sigma_{\nu - d}}/{d E_\nu}$ is the differential neutrino-deuterium cross section~\cite{Kubodera:1993rk} and integrals are computed from 2.2~MeV up to the end point of the $^8$B spectrum.

\par To the above Solar data sets, we add mock data from a Ge dark matter detection experiment. For the general case of a CNS detection at a dark matter detector, the  energy spectrum is 
\begin{equation}
\frac{dR}{dT_R} = \mathscr{N}\int_{E^{\rm min}_\nu} f_\nu(E_\nu)  \left[1- P_{es}(E_\nu) \right]
\frac{d\sigma_{CNS}}{dT_R} dE_\nu
\label{eq:cns}
\end{equation}
where $E^{\rm min}_\nu$ is the minimum neutrino energy required to produce a nuclear recoil of energy $T_R$, and $\mathscr{N}$ is the number of target nuclei per unit of mass of detector material. Dividing by the SSM prediction with $P_{es}(E_\nu) = 0$ gives a prediction in terms of $ f_{8B}$, similar to Equations~\ref{eq:es} and~\ref{eq:rnc}. 

\par A departure from the theoretical predictions of the $^8$B CNS induced nuclear recoil event rate further away from its uncertainty could be interpreted as an evidence for active-to-sterile neutrino oscillation, {\it i.e.} $P_{es}(E_\nu) \neq 0$. It is however worth mentioning that such departures could also be due to non-standard interactions (NSI) \cite{Friedland:2004pp} or from mis-estimation of $\sin^2\theta_w$ at low transferred momentum which has yet to be measured. Combining Solar with reactor, radiogenic, and/or beam CNS measurements would ultimately be required to further assess the validity of a possible evidence of active-to-sterile neutrino oscillation in the Solar sector from CNS measurements~\cite{Anderson:2012pn,Formaggio:2012}. For the remainder of this paper, we will therefore consider that there is no NSI and that the weak charge is perfectly well known. Note that any uncertainties in the weak charge would be quadratically added to the neutrino flux normalization uncertainty.

\par In addition to the mock CNS data, we add mock neutrino-electron elastic scattering for dark matter detectors. For the mock ES data sets we considered {\it pp}, {\it ${}^7Be$}, and {\it CNO} neutrinos using Eq.~\ref{eq:es}, though the dominant contribution comes from {\it pp} neutrinos as illustrated in Fig~\ref{fig:spectrum}. Motivated by the current measurements of the neutrino survival probability in the vacuum dominated regime~\cite{Collaboration:2011nga}, we take the survival probably at {\it pp} neutrino energy to be a constant of $P_{ee} = 0.55$ over the energy range of these three neutrino sources. As for the mock CNS data we consider a detector with perfect energy resolution, $G(T_e,T_{\mathit{eff}}) = 1$. 

\subsection{Likelihood analysis} 

\par Given the above data sets, we are now in position to determine the theoretical parameters that we marginalize over. We take as our set of theoretical parameters $\vec a \equiv \{f_\imath, \sin^2 \theta_{12}, \sin^2 \theta_{13}, \sin^2 \theta_{14}, \Delta m_{21}^2 \}$, where again $\imath = {}^8B, {}^7Be, CNO, pep, pp, hep$. As discussed above, we take a gaussian prior on $\Delta m_{21}^2$ to account for the uncertainty on its measurement from a two flavor analysis. For theoretical priors on the flux normalizations, we take the high metallicity GS98-SFII SSM~\cite{Grevesse:1998bj}-- we note that the constraints on the flux normalizations are unaffected if we were to instead use a low metallicity SSM~\cite{Asplund:2009fu}. Table~\ref{tab:ssm} lists the GS98-SFII SSM priors on the flux normalizations, and Table~\ref{tab:constraints} lists the parameters that we use, the respective observables, and the best constrained parameters from each experiment. Note that we do not include $\Delta m_{41}^2$ as a parameter, because the reactor and Solar data are not sensitive to this mass splitting if it is $\sim 1$ eV$^2$. 

\begin{figure*}
\begin{center}
\begin{tabular}{cccc}
{\resizebox{5.7cm}{!}{\includegraphics{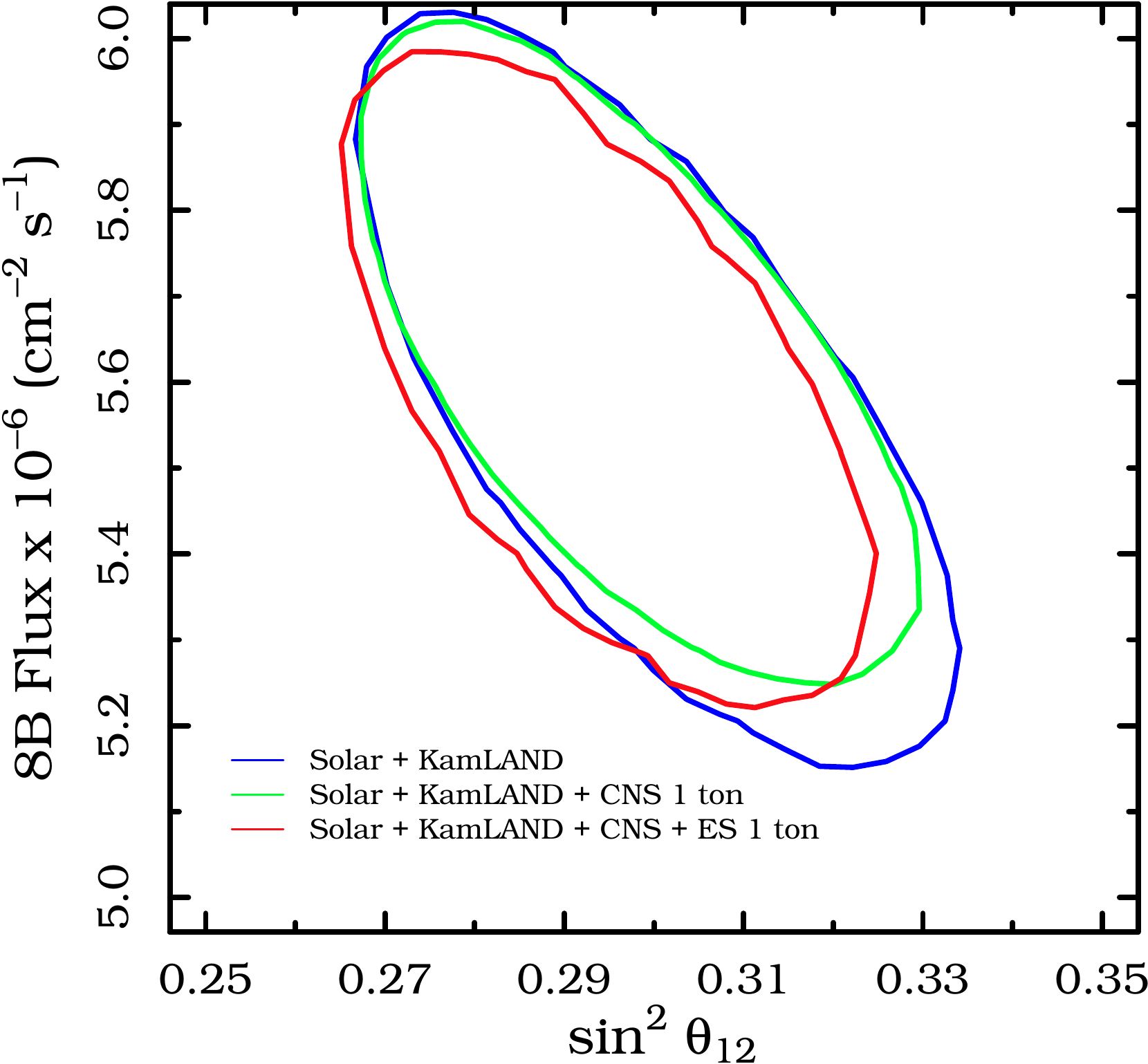}}} &
{\resizebox{5.7cm}{!}{\includegraphics{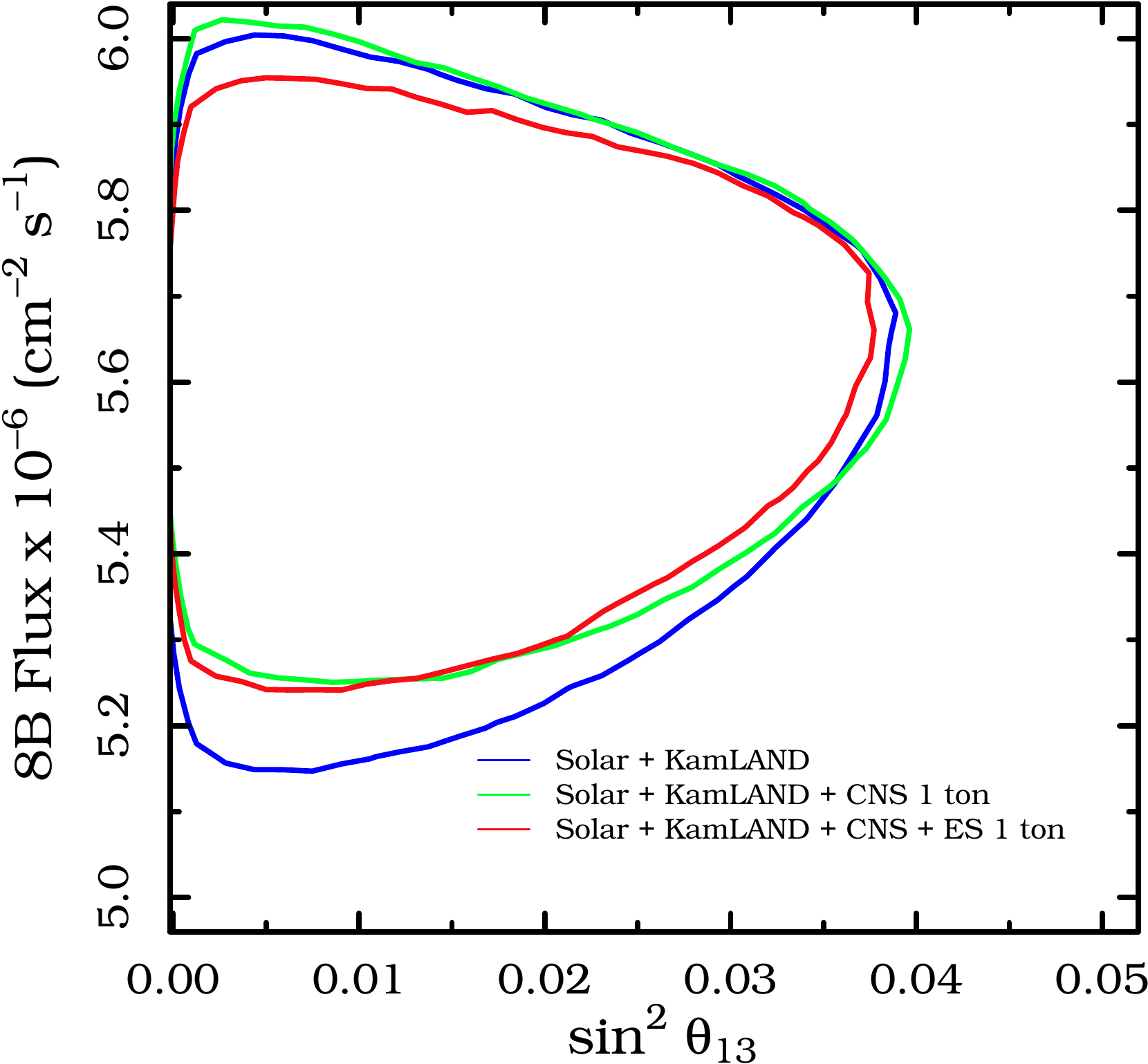}}} &  
{\resizebox{5.7cm}{!}{\includegraphics{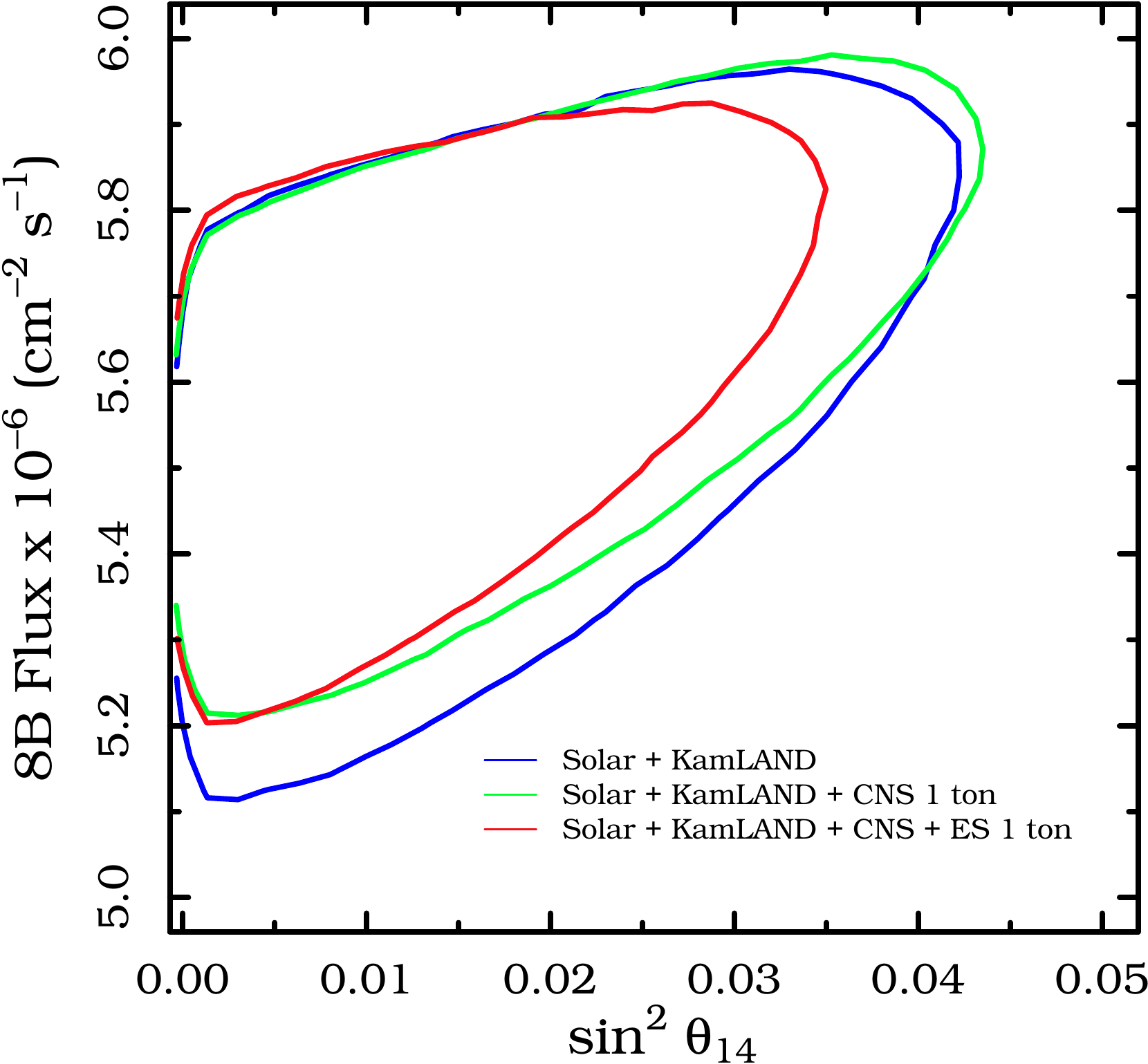}}} \\
{\resizebox{5.7cm}{!}{\includegraphics{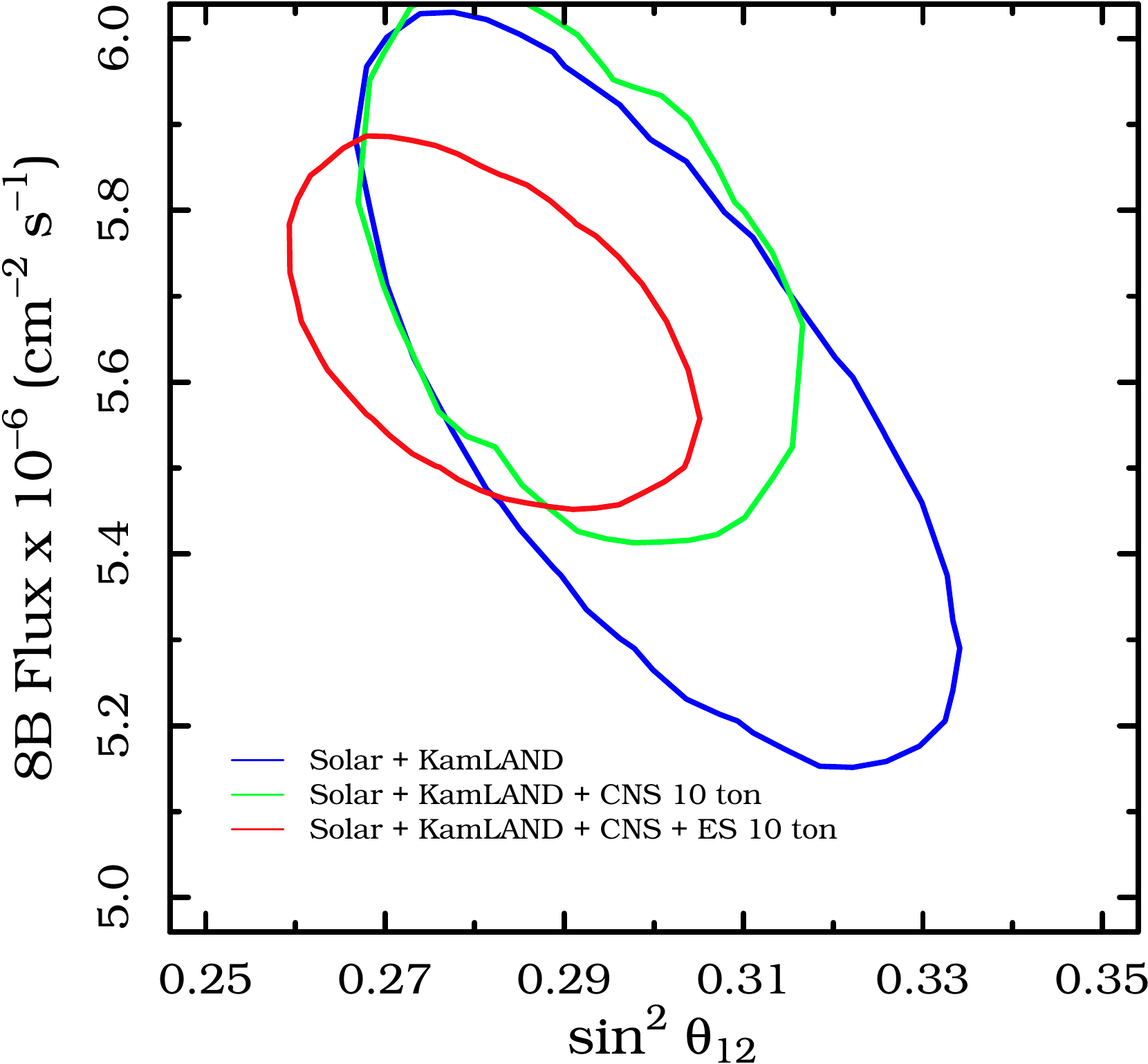}}} &
{\resizebox{5.7cm}{!}{\includegraphics{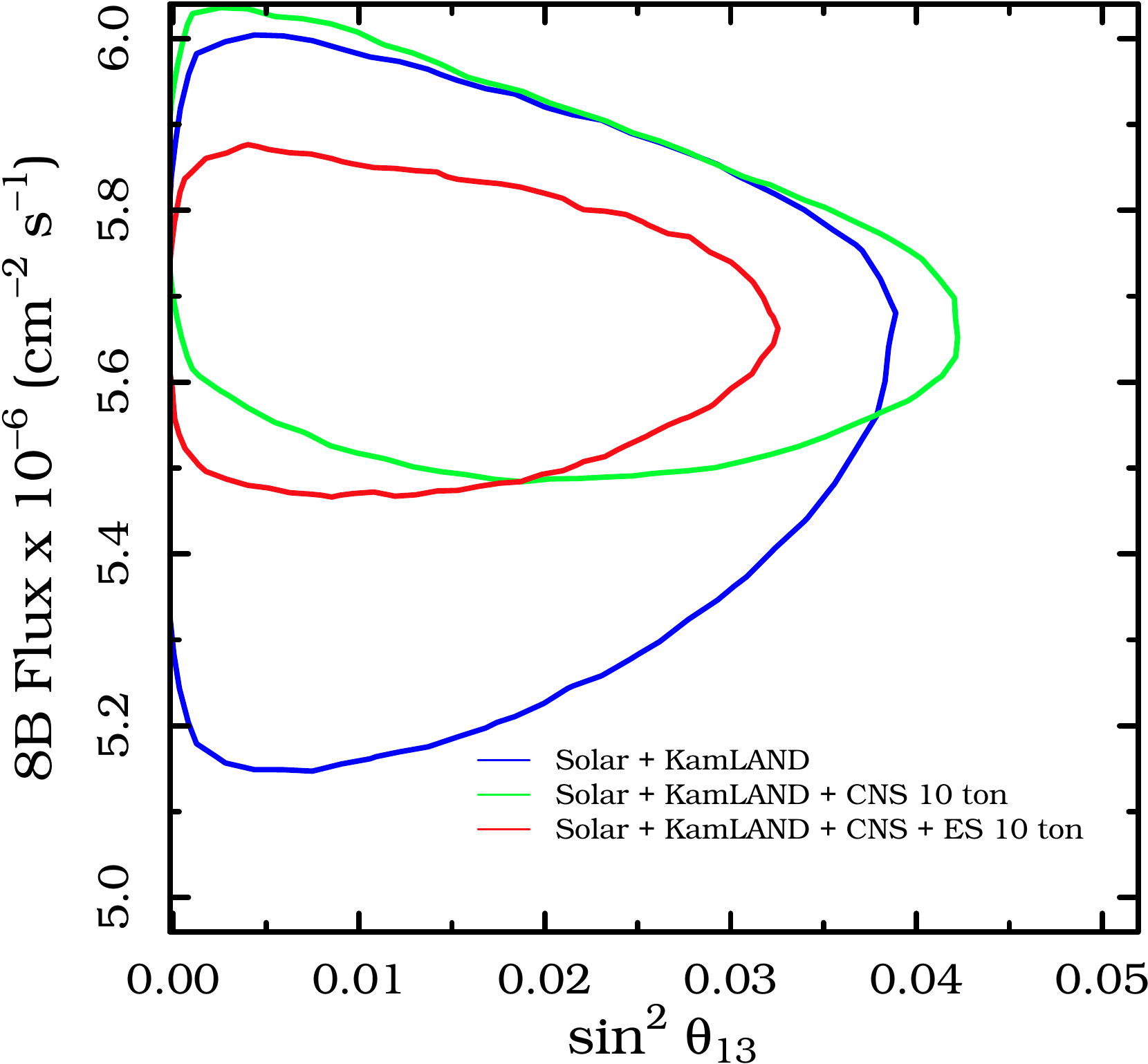}}} & 
{\resizebox{5.7cm}{!}{\includegraphics{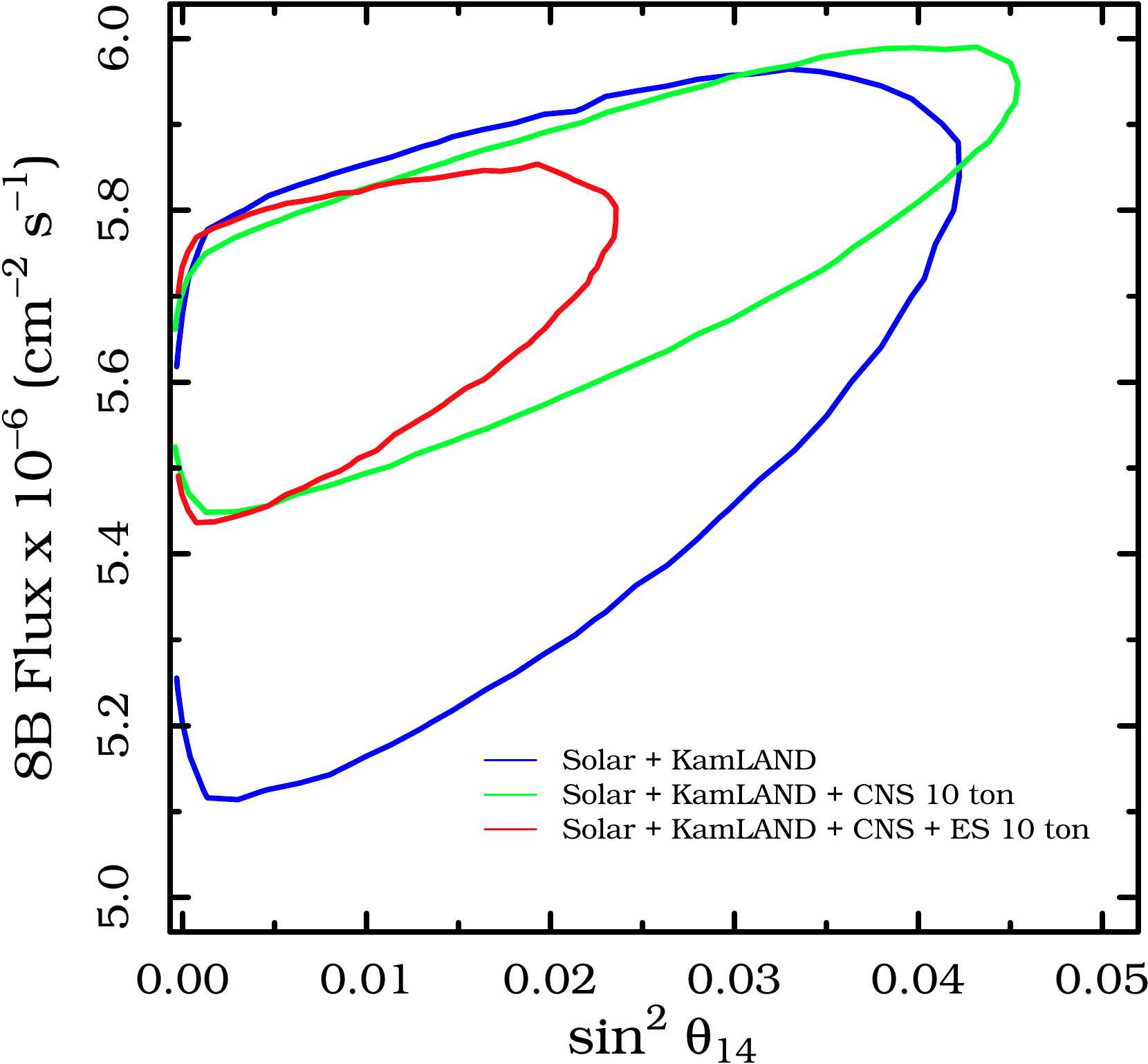}}} \\ 
\end{tabular}
\end{center}
\caption{Derived 90\% C.L.  contours from our MCMC analyses for the normalization of the ${}^8$B flux versus the solar mixing angle $\sin^2\theta_{12}$ (left), the active mixing angle $\sin^2\theta_{13}$ (middle) and the sterile mixing angle $\sin^2\theta_{14}$ (right), when combining current Solar and KamLAND data with future CNS and ES data from a dark matter detector. The top (bottom) panels assume a 1 (10) ton-yr exposure for a Ge detector with a 0.1 keV threshold. These panels highlight the improvement in the measurement of the normalization of the ${}^8$B flux and on the estimation of the neutrino mixing angles with the addition of CNS and ES data from a dark matter detector.}
\label{fig:2Dcontours}
\end{figure*}

\par For a given point in our model parameter space, we use Eq.~\ref{eq:x_kamland},~\ref{eq:es}, ~\ref{eq:rnc}, and~\ref{eq:cns} to determine the theoretical predictions for the different event rates and compare these to the corresponding data sets. To constrain the parameters $\vec a$ we perform a Bayesian analysis in a similar fashion as what has been done in prior solar neutrino analyses~\cite{GonzalezGarcia:2009ya}. We assume a likelihood function of the form ${\cal L} \propto e^{-\chi^2/2}$, with
\begin{equation}
\chi^2_{\rm tot} = \sum_\imath \sum_\jmath \frac{(R_{th,\imath \jmath}(\vec a) -R _{\imath \jmath})^2}{\sigma_\imath^2} + \chi^2_{\rm prior}.
\label{eq:chi2}
\end{equation}
Here $R_{th,\imath \jmath}(\vec a)$ is the theoretical prediction for the rate as a function of the parameters $\vec a$ from the $\imath^{th}$ experiment in the $\jmath^{th}$ energy bin, and $R_{\imath \jmath}$ is given by the rate in an energy bin from one of the aforementioned data sets. In this notation, for the case of an  experiment with one energy bin such as SNO we simply have $\jmath = 1$. Finally, $\chi^2_{\rm prior}$ corresponds to the priors on $f_\imath$ and $\sin^2\theta_{12}$ taken as gaussian distributions as described above.

\par To determine the posterior probability density distributions of the parameters $\vec a$ from the experimental data sets, we utilize a Markov Chain Monte Carlo (MCMC) approach based on the standard metropolis hastings algorithm with a multivariate gaussian proposal function. In order to deal only with independent MCMC samples, we performed a subsampling of the chain to account for both the burn-in and the correlation lengths~\cite{Billard:2010jh}. Using a multivariate gaussian as a proposal function, for all MCMC analyses presented hereafter, we obtained a correlation length around 80, leading to a total of independent samples used for PDF estimations of about 200,000.

%%%%%%%%%%%%%%%%%%%%%%%%%%%%%%%%%%

\section{Results}
\label{sec:results}
Now that our theoretical model and analysis methodology have been discussed in the previous sections, we are in position to first apply our analysis technique using current Solar and KamLAND data. We then move on to study the impact of CNS and ES measurements from dark matter detectors on our understanding of Solar neutrinos. 

\subsection{Solar + Kamland}

\par In order to compare our analysis technique with previous results~\cite{GonzalezGarcia:2009ya,Palazzo:2011rj,Palazzo:2012yf}, we first analyze the Solar and KamLAND data. Figure~\ref{fig:mcmc} shows both the resulting posterior probability densities and the 2D joint distributions for some of our model parameters considering only the current solar and KamLAND data as listed in Table~\ref{tab:constraints}. We again reiterate that in this figure, and in the figures below, we focus on the high metallicity SSM~\cite{Grevesse:1998bj}. Correlations are clearly evident between the mixing angles, in particular between $\sin^2 \theta_{12}$ and $\sin^2 \theta_{14}$. Interestingly one can see that most of the neutrino model parameters exhibit correlations with $f_{8B}$, suggesting that a better measurement of the $^8$B neutrino flux could improve our estimation of the neutrino mixing angles. The anti-correlation between $\sin^2 \theta_{12}$ and $\sin^2 \theta_{14}$ is driven by the KamLAND data, since large values of both of these parameters imply a depleted measured flux from reactors. The anti-correlation between $\sin^2 \theta_{12}$ and $f_{8B}$ is largely driven by the Solar data, in particular the SK measurement of the Solar electron neutrino flux, and its measurement of the mu/tau neutrino flux with a reduced sensitivity. The positive correlation between $\sin^2 \theta_{14}$ and $f_{8B}$ is largely due to the CNS and SNO measurements of the total NC Solar flux. We find that $\sin^2 \theta_{13}$ is largely uncorrelated with any other parameter. 

\par Very generally, we find that the constraints on the parameters deduced from our MCMC analysis are in excellent agreement with previous determinations of these parameters. These results are summarized in Table~\ref{tab:params}. The upper limit that is deduced from the posterior probability density of $\sin^2 \theta_{14} < 0.034$ (at 90\% C.L.) is in good agreement with the upper bounds quoted in Refs.~\cite{Palazzo:2011rj,Palazzo:2012yf}. Also, the constraints on $\Delta m_{21}^2$ and $f_{8B}$ are consistent with the input priors, and our measurement of $\sin^2 \theta_{12}$ is consistent with previous results, even though we have a flat prior on this quantity. It is worth emphasizing that the goal of this paper is not to perform a perfectly complete and detailed 3+1 analysis but rather to show, for the first time, what a dark matter detector could bring to the field of neutrino physics within the scope of a simplified 3+1 analysis, as presented in Sec.~\ref{sec:models}. 

\subsection{Including data from a low-threshold dark matter detector}

\begin{figure*}%[htp]
\begin{center}
\includegraphics[width=0.49\textwidth]{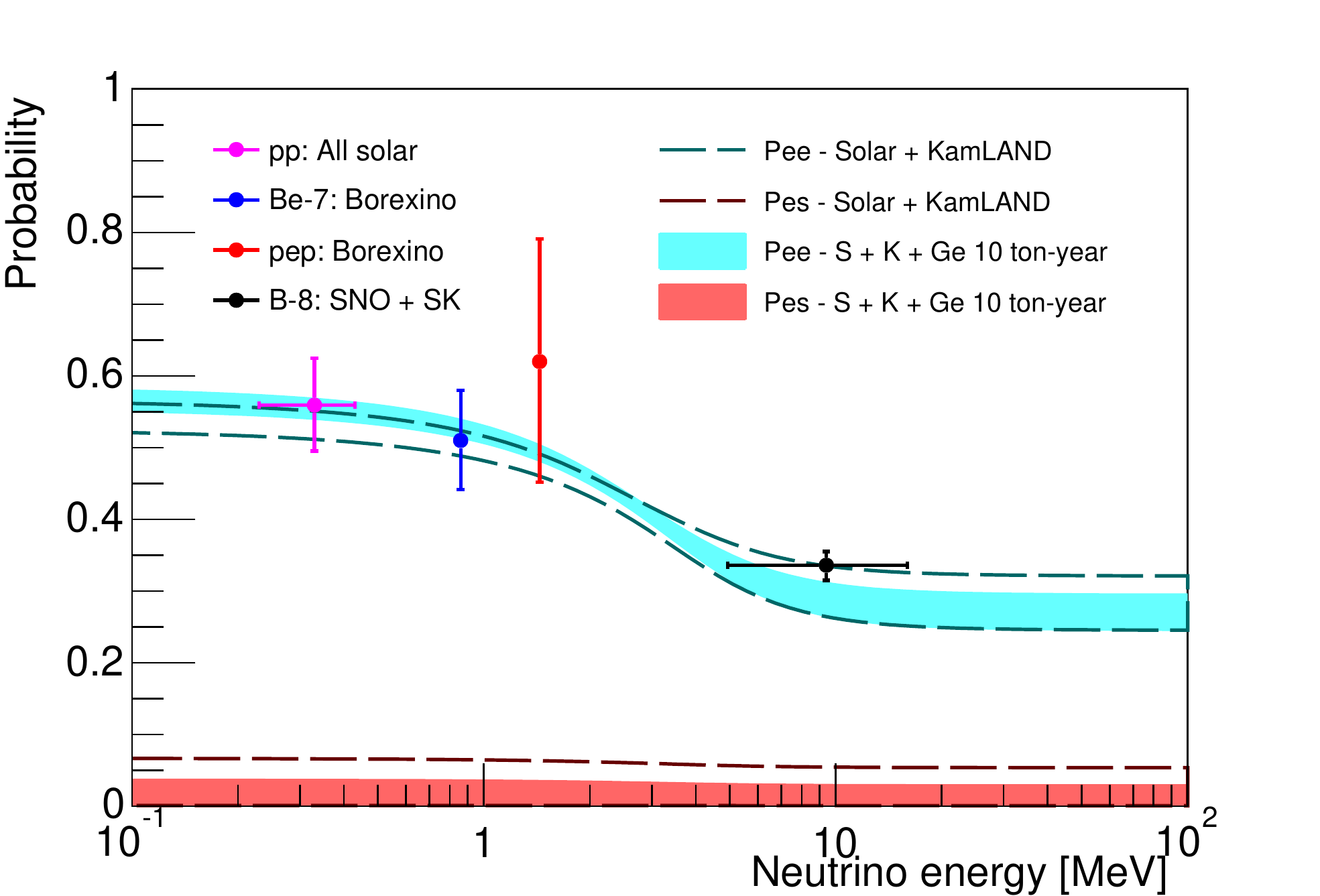} 
\includegraphics[width=0.49\textwidth]{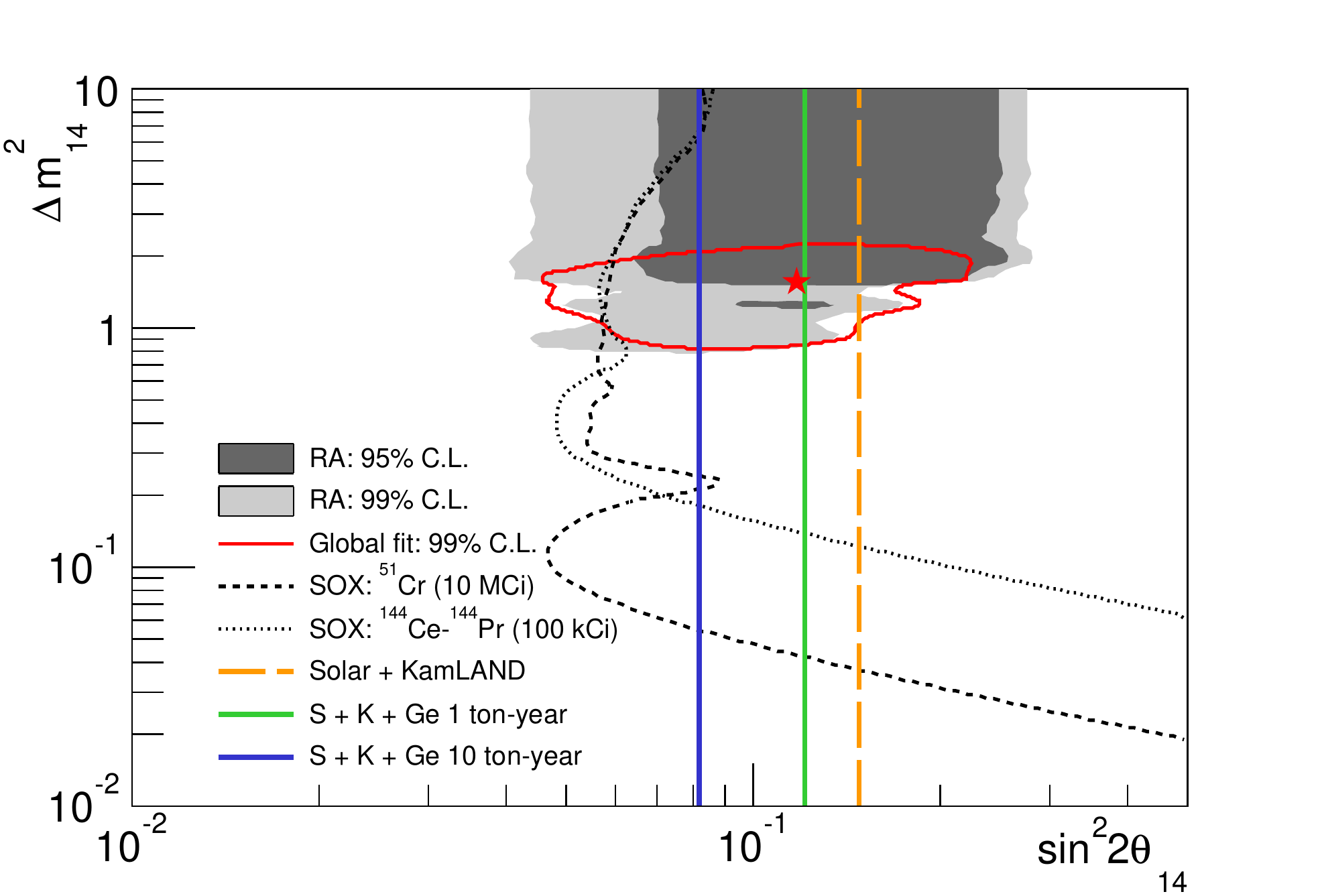} 
\end{center}
\caption{{\it Left:} Contours at 95\% C.L. on the electron neutrino survival probability $P_{ee}$ (cyan) and transition probability into a sterile neutrino $P_{es}$ (red) as a function of the neutrino energy. The two set of bands correspond to the case Solar + KamLAND (dashed lines) and to the case Solar + KamLAND + CNS + ES with a 10 ton-year exposure (filled contours). The contours are determined from Bayesian marginalization of the previously discussed MCMC analyses. Also shown are the current constraints on the neutrino-electron survival probability derived assuming no existence of sterile neutrinos~\cite{Bellini:2013lnn}.
{\it Right:} Projected limits on the active-to-sterile mixing angle $\sin^2 \theta_{14} \equiv \sin^2 \theta_{ee}$ using all current Solar and KamLAND data plus a 1 (green) and 10 (blue) ton-year exposure of a Ge dark matter detector sensitive to both CNS and ES neutrino induced events. The highlighted regions are the favored solutions for the reactor anomaly at the 95\% and 99\% C.L.~\cite{Giunti:2012tn}. The red contour corresponds to the 99\% C.L. constraint and best fit point derived from a global analysis of both neutrino disappearance and appearance data~\cite{Giunti:2013aea}. The dashed grey curves are the projected limit from the SOX experiment~\cite{D'Angelo:2014vgk,Borexino:2013xxa}.
\label{fig:probs}
}
\end{figure*}

In this section, we estimate how a low-threshold dark matter detector with a ton-scale exposure could improve on the results presented in Figure~\ref{fig:probs}. As discussed above, such an experiment should give the unique opportunity to probe the solar neutrino sector at both low and high energies, {\it i.e.} in the vacuum and matter dominated regimes. To do so, we have added simulated data (CNS + ES) to the previously described MCMC analysis using current data from other experiments listed in Table~\ref{tab:constraints}. We have simulated data from the theoretical CNS and ES event rate spectra, as shown in Fig.~\ref{fig:spectrum}, in a model independent fashion by considering only current data. As discussed above, for the ES event rate we used the averaged $P_{ee}$ value as derived from the combined analysis of all solar experiments sensitive to $pp$ neutrino (see pink dot in left panel of Fig.~\ref{fig:probs}) which were derived with no sterile neutrinos. The CNS data were generated considering $\sin^2\theta_{14} = 0$, {\it i.e.}  assuming no active-to-sterile transition.

\par Figure~\ref{fig:2Dcontours} shows how constraints at 90\% C.L. on selected parameters evolve with the different data sets considered: Solar + KamLAND (blue), Solar + KamLAND + CNS (green), and Solar + KamLAND + CNS + ES data from a dark matter detector (red). We considered exposures of 1 (top panels) and 10 (bottom panels) ton-year. For the Ge dark matter detector, we binned the data from 0.1~keV to 100~keV with 10 (20) bins for the 1 (10) ton-year exposure. 

\par In general we find that the most substantial improvement by including CNS at dark matter detector is in the determination of $f_{8B}$, {\it i.e.} the $^8$B neutrino flux normalization. For example with the addition of CNS data from a Ge dark matter detector with an exposure of 1 (10) ton-year to existing solar and KamLAND data, we find that $f_{8B}$ is determined with a precision of 3.2\% (2.2\%). With this level of uncertainty, the addition of CNS data alone will be able to clearly distinguish between the high metallicity GS98-SFII~\cite{Grevesse:1998bj} and low metallicity AGSS09-SFII~\cite{Asplund:2009fu} SSMs, which have respective flux normalizations and theoretical uncertainties of $5.58 \times 10^6 (1 \pm 0.14)$ cm$^{-2}$ s$^{-1}$ and $4.59 \times 10^6 (1\pm 0.14)$ cm$^{-2}$ s$^{-1}$.  

\par With $f_{8B}$ constrained by the CNS data, the addition of ES data from a dark matter detector then improves the constraints on $\sin^2\theta_{14}$. The constraints on $\sin^2\theta_{14}$ are most substantially improved when moving from a 1 ton-year to 10 ton-year exposure. It is additionally worth noting that due to the different correlations between the neutrino flux normalizations and the neutrino mixing angles, a CNS and ES measurement from a dark matter detector combined with reactor and other solar experiments can still substantially improve on the neutrino parameters. This is indeed illustrated in Fig.~\ref{fig:2Dcontours} where we show the derived constraints in the ($f_{8B}$, $\sin^2\theta_{12}$) plane. Such a result suggests that CNS and ES at dark matter detectors, combined with existing experiments, can improve our estimates of the different active-to-active oscillations as a function of the neutrino energy in the context of a given neutrino model (3+1 in this case). It is also worth noticing that in the case of the Solar + KamLAND + CNS + ES analysis with a 10 ton-year exposure, the reconstructed value of  $\sin^2\theta_{12}$ is slightly shifted to lower values compared to the other analyses presented in Fig.~\ref{fig:2Dcontours}. This is because we generated our mock ES data using $P_{ee} = 0.55$ for the {\it pp} neutrinos as motivated by current measurements (see the pink dot in Fig.~\ref{fig:probs} left panel) and not from a global analysis that tends to favor lower values of $P_{ee}$, as derived from our Solar + KamLAND combined anaysis. This leads to a lower reconstructed value of $\sin^2\theta_{12}$ compared to other analyses presented in Fig.~\ref{fig:2Dcontours}. We checked that the conclusions of our work are fairly insensitive to the particular choice of input value of $P_{ee}$ at $pp$ neutrinos and that the interest here is to quantify how much the uncertainties on the solar neutrino physics parameters can be reduced with the addition of a dark matter experiments to the current Solar neutrino data.

\par From the posterior probability densities of the considered parameters in our MCMC analysis we can also determine the shape of the transition and survival probabilities as a function of neutrino energy. Figure~\ref{fig:probs} shows the derived 95\% C.L. bands on the neutrino-electron survival probability in cyan and the neutrino-electron to sterile neutrino transition probability in red. The dashed lines correspond to the Solar + KamLAND case while the filled contours are after the inclusion of a 10 ton-year low-threshold Ge detector. Note that the filled contours with the Ge data are shifted relative to the dashed contours with Solar + KamLAND data only; again and as discussed above, this is because of our assumption of a constant electron neutrino survival probability of $P_{ee} = 0.55$ for $pp$ neutrinos, as suggested by current experimental measurements. We see that regarding the overall uncertainties, as more Solar neutrino data sets are added to the KamLAND data, both $P_{ee}(E_\nu)$ and $P_{es}(E_\nu)$ become more strongly constrained, by about 50\%. 

\par Indeed, by measuring both neutrino-electron scattering at low energies, from $pp$ neutrinos around 0.4 MeV, and coherent neutrino scattering, from $^8$B neutrinos around 10 MeV, a dark matter detector has the unique opportunity to study neutrino physics within both the vacuum and the matter dominated regime. This is of particular interest as the exact shape of the transition, happening around 2 MeV, can be influenced by the existence of sterile neutrinos and/or non standard interactions~\cite{Friedland:2004pp} to which a dark matter detector would then be sensitive to. In all cases, we can clearly see that the active-to-sterile neutrino oscillation is fairly constant as a function of the neutrino energy. Interestingly, measuring CNS with $^8$B neutrinos will allow future low-threshold dark matter experiments to also place an upper bound on the averaged $P_{es}$ transition probability in a model independent fashion. However, such approach would require significant reduction of the theoretical uncertainty on the $^8$B neutrino flux which is about 14\%~\cite{reviews}.\\

\par A dark matter detector can also place interesting constraints on the active-to-sterile neutrino oscillations related to $\sin^2\theta_{14}$. Indeed, the right panel of Fig.~\ref{fig:probs} shows how our projected limits from a 1 (10) ton-year Ge detector in green (blue) compares to other current and projected measurements of the active to sterile mixing angle. This figure indicates that a 1 ton-year experiment could reach the best fit point of the global analysis from~\cite{Giunti:2013aea} and that a 10 ton-year Ge detector will effectively probe most of the parameter space that can explain the reactor anomaly~\cite{Giunti:2012tn}. It also shows that upcoming dark matter experiments could be competitive with the expected sensitivity of the forthcoming SOX experiment~\cite{Borexino:2013xxa,D'Angelo:2014vgk}. Therefore, the Solar neutrino measurements with a dark matter detector sensitive to both ES and CNS that we have discussed in this paper can be complementary to experiments that are planned to probe active to sterile oscillations in the Solar sector~\cite{Borexino:2013xxa,Djurcic:2013oaa}. 

\section{Discussion and Conclusions} 
\label{sec:conclusion}
\par We have discussed the implications of the measurement of Solar neutrinos in dark matter detectors through both the coherent neutrino-nucleus scattering channel and the neutrino-electron elastic scattering channel. Most generally, our results show that a CNS detection of ${}^8$B neutrinos will provide a measurement of the ${}^8$B flux normalization to a few percent, and most importantly will provide an independent test of high and low metalicity Solar models. For a 10~ton-year detector, we found that a measurement of elastic scattering $pp$ neutrinos will help reducing the uncertainty on the neutrino mixing parameters which are mostly relevant to the vacuum dominated regime. Furthermore, we show that combining the ES and CNS measurements will further improve on  both the estimation of the neutrino electron survival probability over all energies and the sensitivity to the sterile neutrino mixing angle by about a factor of 2 within a 3+1 neutrino model. This implies that dark matter detectors are uniquely positioned to study both the high and low energy survival probability simultaneously through two distinct channels and allow for a competitive and alternative way to probe the possible existence of sterile neutrinos as hinted by the reactor anomalies.

\par The analysis in this paper has primarily focused on Solar and KamLAND data. It is also possible to consider data sets that better constrain some of the parameters that we have discussed. As an example we have not included short baseline data from the Daya Bay~\cite{An:2012eh}, RENO~\cite{Ahn:2012nd}, and Double Chooz~\cite{Abe:2012tg} reactor experiments which have recently measured a non-zero value of $\sin^2 \theta_{13}$. Though a detailed inclusion of these data sets is beyond the scope of our simplified analysis, it is possible to obtain an estimate of what the non-zero $\sin^2 \theta_{13}$ measurements imply for our results. Indeed, considering a simplified analysis of the ratio of the observed event rates  in the near and far detectors from Daya Bay and RENO we  found that the overall sensitivity to $\sin^2 \theta_{14}$ can be improved by a factor of 2.

\par While our 3+1 analysis focused on a model with a mass splitting $\Delta m^2 \sim $ eV$^2$ relative to the other active neutrinos, it is important to recognize that our results are more broadly applicable to models with much different mass splittings. In the future it will be interesting to consider for example the impact of sterile neutrinos with smaller mass splitting than we have considered here~\cite{deHolanda:2010am} and also include the possibility of non-standard neutrino interactions~\cite{Palazzo:2011vg,Bonventre:2013loa}. Combining Solar neutrino data from a dark matter detector with present neutrino data sets should lead to more interesting constraints on these and other theories of extended neutrino sectors. \\

{\bf Acknowledgments}-- We thank John Beacom, Rafael Lang, and Antonio Palazzo for discussions. LES was supported by NSF grant PHY-1417457, and EFF and JB by grant NSF-0847342. This material is based upon work supported by the National Science Foundation under Grant No. CNS-0723054.

\end{document}